\documentclass[aps,prl,reprint,superscriptaddress]{revtex4-1}

\usepackage{amsmath}
\usepackage{xcolor}
\usepackage{graphicx}
\usepackage{epstopdf}
\usepackage{hyperref}

\begin{document}

\title{Diffusing Up the Hill: Dynamics and Equipartition in Highly Unstable Systems}

\author{Martin \v{S}iler}
\email[]{siler@isibrno.cz}
\affiliation{Institute of Scientific Instruments of the Czech Academy of Sciences,  Kr\'{a}lovopolsk\'{a} 147, 612 64 Brno, Czech Republic}
\author{Luca Ornigotti}
\affiliation{Department of Optics, Palack\' y University, 17. listopadu 1192/12,  771~46 Olomouc, Czech Republic}
\author{Oto Brzobohat\'y}
\affiliation{Institute of Scientific Instruments of the Czech Academy of Sciences,  Kr\'{a}lovopolsk\'{a} 147, 612 64 Brno, Czech Republic}
\author{Petr J\'akl}
\affiliation{Institute of Scientific Instruments of the Czech Academy of Sciences,  Kr\'{a}lovopolsk\'{a} 147, 612 64 Brno, Czech Republic}
\author{Artem~Ryabov}
\email[]{rjabov.a@gmail.com}
\affiliation{Charles University, Faculty of Mathematics and Physics, Department of Macromolecular Physics, V Hole\v{s}ovi\v{c}k\'ach 2, 180 00 Praha 8, Czech Republic}
\author{Viktor Holubec}
\affiliation{Charles University, Faculty of Mathematics and Physics, Department of Macromolecular Physics, V Hole\v{s}ovi\v{c}k\'ach 2, 180 00 Praha 8, Czech Republic}
\affiliation{Universit{\" a}t Leipzig, Institut f{\" u}r Theoretische Physik, Postfach 100 920, D-04009 Leipzig, Germany}
\author{Pavel Zem\'{a}nek}
\affiliation{Institute of Scientific Instruments of the Czech Academy of Sciences,  Kr\'{a}lovopolsk\'{a} 147, 612 64 Brno, Czech Republic}
\author{Radim Filip}
\email[]{filip@optics.upol.cz} 
\affiliation{Department of Optics, Palack\' y University, 17. listopadu 1192/12,  771~46 Olomouc, Czech Republic}

\date{\today}
\begin{abstract}
Stochastic motion of particles in a highly unstable potential generates a number of diverging trajectories leading to undefined statistical moments of the particle position. This makes experiments challenging and breaks down a standard statistical analysis of unstable mechanical processes and their applications. A newly proposed approach takes advantage of the local characteristics of the most probable particle motion instead of the divergent averages. We experimentally verify its theoretical predictions for a Brownian particle moving near an inflection in a highly unstable cubic optical potential. The most-likely position of the particle atypically shifts against the force despite the trajectories diverge in the opposite direction. The local uncertainty around the most-likely position saturates even for strong diffusion and enables well-resolved position detection. Remarkably, the measured particle distribution quickly converges to the quasi-stationary one with the same atypical shift for different initial particle positions. The demonstrated experimental confirmation of the theoretical predictions approves the utility of local characteristics for highly unstable systems which can be exploited in thermodynamic processes to uncover energetics of unstable systems.  
\end{abstract}


\maketitle

\section{Introduction} 
Unstable stochastic dynamics of mechanical objects is a common ingredient of processes inside mechanical machines. They use explosive fuel to move a piston or its microscopic equivalent ATP to perform individual strokes \cite{Vale88,Schliwa2003, AstumianFarDis2016, ErbasCakmak2015}. However, instability generates rapidly diverging trajectories which complicate the description of motion, its experimental observation and applications. Such trajectories can make all statistical averages increasing very fast or even diverging. The standard deviation of position can diverge faster than its mean hence the observed average motion quickly becomes uncertain. Moreover, the probability density of position can develop a heavy tail and, therefore, all its moments will diverge  \cite{FilipJOpt16, OrnigottiPRE18}. All this limits statistical description of transient effects in unstable potentials and makes experimental observations challenging. The detrimental effects appear even in a strongly overdamped regime, where a system intensively dissipates energy to the environment. The simplest example is an overdamped Brownian particle diffusing in the highly unstable cubic potential $V(x) = \mu x^3/3$. The Langevin equation for the particle position reads 
\begin{equation}
\gamma \frac{dx}{dt}= - V'(x(t))  + \sqrt{2 \gamma k_B T}\xi (t),
\label{eq:LE}
\end{equation}
where $\gamma$ is the friction coefficient, $T$ is the ambient temperature, $k_B$ the Boltzmann constant, and $\xi(t)$ stands for the delta-correlated Gaussian white noise, $\langle \xi(t) \rangle = 0$, $\langle \xi(t)\xi(t')\rangle=\delta(t-t')$.

Eq.~\eqref{eq:LE} with the cubic potential $V(x) = \mu x^3/3$ models dynamics at the saddle-node bifurcation \cite{Sigeti1989, ReimannBroeck94}, which arises in several nonlinear stochastic models of physics, biology and chemistry. Examples include optical bistability in lasers \cite{SanchoPRA1989, SanchoPRA1991, ArecchiPRA1971}, firing of neurons \cite{Lindner2003, Brunel2003}, Brownian ratchets~\cite{ReimannPRL01, ReimannPRE02, Dean1, PavelPRL2017}, or nonlinear maps~\cite{HirschPRA1982}.   
Even though we primarily focus on the cubic nonlinearity with the inflection point, the described methodology is broadly applicable to any highly unstable potential. This we demonstrate in the Supplemental Material~\cite{SI} (SM) on prominent examples: the unstable potential with no stationary point, and the unstable potential with a local minimum. The latter is frequently used to study decays from metastable states in condensed matter models~\cite{TretiakovPRB2003, SpagnoloEntropy2017}.

The instability in the cubic potential  has been theoretically analyzed by means of statistical moments for times shorter than appearance of any diverging trajectory \cite{FilipJOpt16} and using first passage times for distances far away from the instability \cite{Arecchi1982, YoungPRA1985, SanchoPRA1989, HirschPRA1982, Sigeti1989, SanchoPRA1991, Caceres1995, MantegnaPRL1996, AgudovPRE1998, Lindner2003, Brunel2003, FiasconaroPRE2005, CaceresJSP2008, RyabovPRE16,  HanggiRevModPhys1990}. In both cases, dynamical effects have been found to be stimulated by the initial temperature of the particle and the temperature of the surroundings. Recently, both described regimes have been observed experimentally for a micron-sized particle trapped in optical tweezers~\cite{SilerSR17}. However, the transient regime beyond short-time approximation remained without any description and physical understanding. The lack of description and understanding undermines applications of unstable stochastic dynamics in nanotechnology and in quantum technology.

Recently, a new methodology focused on the most-likely position of the particle in unstable potential has been developed \cite{OrnigottiPRE18}. It uses directly measurable local features of the probability density of position instead of global diverging statistical moments. During the transient dynamics, this methodology evaluates a shift of the probability density maximum instead of the mean value, and local curvature around the maximum instead of the standard deviation.
Moreover, a transient decay of the probability density function of particle position (PDF) at late times follows $P(x,t) \sim Q_{\rm st}(x) {\rm e}^{- \lambda_0 t}$, where $Q_{\rm st}(x)$ is a time-independent normalized PDF independent of initial conditions, and $\lambda_0$ is a positive decay rate. The PDF $Q_{\rm st}(x)$ is the so called quasi-stationary distribution \cite{Yaglom1947, bookQSD, PolletURL,  Nasell1995, Hastings2004, Steinsaltz2004, TracerPRE2014}  and it predicts both the shift and local curvature of position PDF in the long-time limit.

\begin{figure}[t]
    \centering
     \includegraphics[width=\columnwidth]{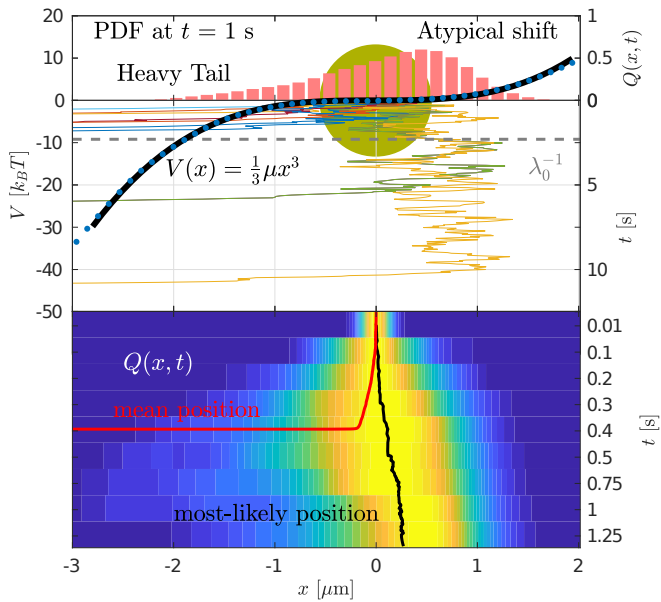}
     \caption{The most-likely motion of a particle in the highly unstable cubic potential. Upper-left axis: the potential $V(x)$ recovered from the experimental trajectories (blue points) fitted by the cubic dependence (black solid line). Upper-right axis: the normalized measured PDF for $x(t$ at $t=1$~s (histogram). Central-right axis: a typical set of 9 measured trajectories starting at $x=0$. The yellow sphere illustrates the size of used colloidal particles (diameter $\approx 1$~$\mu$m).  
Lower panel: individual stripes show a measured PDF at different times (the lower-right axis). 
The mean particle position (the red curve) diverges around $t = 0.3$~s. In fact, all measured moments diverge at the same time. Contrary to this, the black curve demonstrates the regular evolution of the most probable particle position (the position of PDF maximum) shifting to the right from $x=0$ against the direction of acting force. 
}
     \label{fig:il}
\end{figure} 

In this Letter, we experimentally demonstrate the applicability of this local description focused on the most-likely particle position in the unstable cubic potential illustrated in Fig.~\ref{fig:il}. The experimental data agree well with theory even for small sample of measured trajectories, showing that the theory is robust and applicable even under severe experimental imperfections. We unambiguously distinguish the atypical shift of the most-likely position for transients longer than the short-time limit and experimentally verify a fast convergence of the position PDF to the quasi-stationary state. Finally, we derive and experimentally verify the quasi-stationary generalization of the equipartition theorem showing that the quasi-stationary state has high energetic content which can be utilized in a postselection process removing divergent trajectories. 

The experiments verify that the methodology proposed in Ref.~\cite{OrnigottiPRE18} removes existing limitations in description and understanding of transients in highly unstable potentials. Experiments with other typical unstable potentials are included in SM~\cite{SI}. They demonstrate wide applicability of the approach beyond the cases discussed in Ref.~\cite{OrnigottiPRE18}. Direct applications of such most-likely motion in unstable systems are expected in Brownian motors. The methodology can be translated to recently achieved underdamped regime \cite{FonsecaPhysRevLett2016, Ricci2017, Rondin2017} and developing experiments approaching quantum regime \cite{JainPhysRevLett2016, HoangPhysRevLett2016, Rahman2017}.

\section{Experiments}

The quasi-onedimensional cubic potential profile, $V(x) = \mu x^3 /3$,  $\mu = (4.07 \pm 0.03)\ k_B T$ $\mu \mathrm{m}^{-3}$, was created by two pairs of counter-propagating Gaussian laser beams in the configuration that ensures a conservative optical force~\cite{CizmarLPL11, ZemanekJOpt16}. We tracked the particle position with the CCD camera at 2000 fps. The frame rate is fast enough to analyze the transient unstable stochastic dynamics.  Technical details of the setup and data processing are described in SM~\cite{SI}.  

Strong instability of the cubic potential limits the duration of experiments.  As Fig.~\ref{fig:il} illustrates, the number of non-diverging trajectories decreases rapidly with time. 
Thus also the size of ensemble of measured particle positions rapidly decreases, because the divergent trajectories are excluded from the statistics. For example at $t=1$~s we processed into PDFs roughly 55\% of the total number of recorder trajectories (starting at $x_0=0$). At longer times the number of non-diverging trajectories decays exponentially $\sim {\rm e}^{-\lambda_0 t}$. Therefore, due to strength of instability, a sufficient statistics of particle position is practically unreachable at longer times. This behavior is generic for highly unstable potentials (see other examples in the Supplemental Material). 

Furthermore, for such unstable potential the total number of trajectories recorded with a specific particle was rather limited. Typically we performed  $\approx 100$  measurement cycles under the same experimental conditions. After that a fine readjustment of the experimental system was needed to compensate systematic changes like intrinsic mechanical drifts. Consequently, the profile of the optical potential slightly varied from the previous one and thus the new set of trajectories could not be mixed with the previous one into larger statistical ensemble of measured trajectories.

\begin{figure}[t!]
    \centering
    \includegraphics[width=0.98\columnwidth]{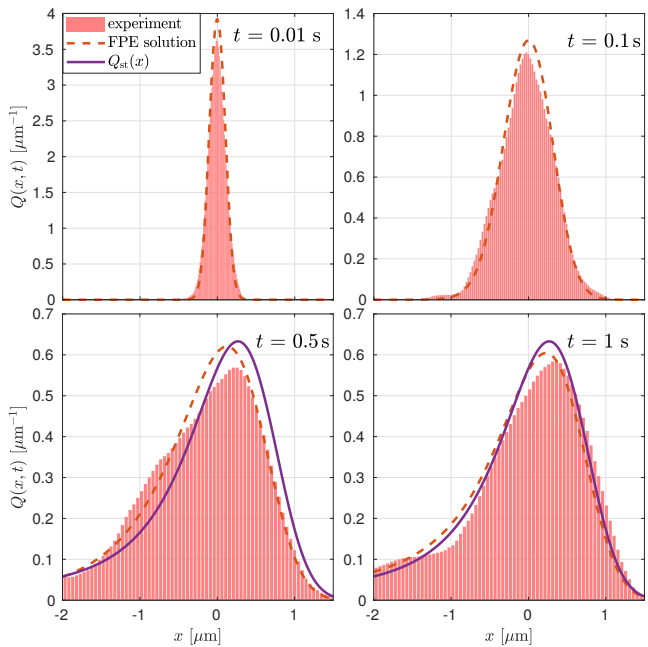}\\
    \caption{Convergence of the position PDF $Q(x,t)$ towards the quasi-stationary state $Q_{\mathrm{st}}(x)$. The experimentally measured PDFs (histograms) are in a reasonable agreement with the numerical solutions of the corresponding Fokker-Planck equation (dashed lines). Solid lines in the lower panels depict the quasi-stationary PDF $Q_{\mathrm{st}}(x)$ calculated from Eq.~(\ref{eq:QstDiff}). The atypical shift of $x_{\rm max}(t)$ right from $x=0$ is clearly observable.} 
    \label{fig:PT} 
\end{figure} 

\section{The most likely position and its atypical shift}

A Brownian particle in the optical cubic potential and results of a typical measurement are illustrated in Fig.~\ref{fig:il}.  The histogram in the figure illustrates measured PDF at time $t=1$~s.  The unstable cubic potential induces three crucial effects in the position PDF: (i) a heavy tail for $x\ll -(3 k_BT/\mu)^{1/3}$, (ii) a light tail for $x\gg (3 k_BT/\mu)^{1/3}$, and (iii) a shift of the PDF maximum away from $x=0$. Moreover, owing to the thermal noise, the PDF after a short time loses all information about the initial particle position and attains a time-independent (quasi-stationary) spatial shape.

\begin{figure}
    \centering
    \includegraphics[width=0.98\columnwidth]{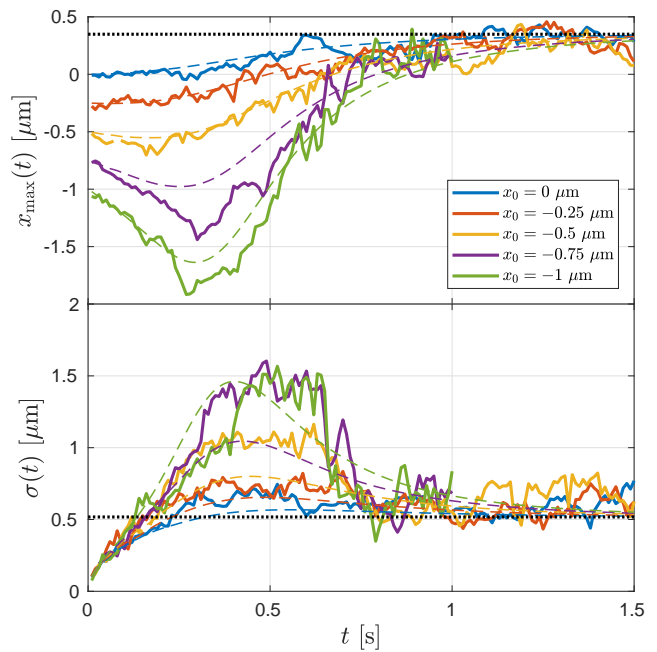}
    \caption{Atypical shift of the PDF maximum against the acting force. The upper panel: The position of PDF maximum $x_{\rm max}(t)$ for different initial positions $x_0$ evaluated from the experimental data (solid lines) and simulations (dashed lines). When $x_0<-0.5$, all trajectories diverge extremely fast due to strong instability. The green and the purple curves stop at time $t=1$~s, because there is not enough datapoints to reconstruct PDFs for $t>1$~s. The lower panel shows $\sigma(t)$ evaluated for the same data. The transient dynamics quickly converges to the quasi-stationary state [constant dotted lines calculated from $Q_{\mathrm{st}}(x)$ defined by Eq.~(\ref{eq:QstDiff})].}
    \label{fig:resx0}
\end{figure} 

The weight of the heavy tail quickly increases with time because of a growing number of diverging trajectories (shown in the middle panel of Fig.~\ref{fig:il}). Several of them diverge in a short time, consequently averages $\langle x(t) \rangle$ (cf.\ red curve in the lower panels) and $\langle x^{2}(t) \rangle$ computed over all measured trajectories quickly grow above all bounds. These average global characteristics are strongly influenced by the divergence and hence bare no meaning for the description of the particle position \cite{SilerSR17}. The heavy tail negatively influences all quantities based on averaging over all trajectories. 

Instead, as proposed in a recent theoretical work \cite{OrnigottiPRE18}, it is beneficial to focus on local characteristics. First of them is the most likely particle position $x_{\mathrm{max}}(t)$ given by the maximum of the PDF $Q(x,t)$, see the black curve in the lower panel of Fig.~\ref{fig:il}. The second one quantifies a local uncertainty around the most probable position. It is defined as the (inverse) normalized curvature at the PDF maximum, 
\begin{equation} 
   \sigma^2(t) = \frac{Q(x_{\mathrm{max}}(t),t)}{\left| \partial^2_{xx} Q(x_{\mathrm{max}}(t),t)\right|}.
   \label{eq:sigma}
\end{equation} 
Different from the diverging mean and variance, the local quantities $x_{\mathrm{max}}(t)$ and $\sigma(t)$ remain finite and attain finite constant values at long times (see Fig.~\ref{fig:resx0} and the discussion below). The ratio $x_{\rm max}(t)/\sigma(t)$ specifies a local visibility of the most-likely particle position.

Measured time evolution of the local quantities is shown in Fig.~\ref{fig:resx0} for different initial positions $x_0$, $x_0=-1$, $-0.75$,  $-0.5$, $-0.25$ and $0$ $\mu$m. For large negative $x_0$ the maximum $x_{\rm max}(t)$ drifts first in the direction of acting force, $-\mu x^2$. Later it passes through its minimal value and starts to shift back {\em against} the force. Eventually, $x_{\rm max}$ becomes positive and independent of time and of the initial position. This is accompanied by a non-monotonous dynamics of the local standard deviation $\sigma(t)$. For large negative $x_0$, it passes through a pronounced maximum and later decreases [which corresponds to a sharper peak of $Q(x,t)$].

\section{Quasi-stationary state in the unstable potential}

A remarkable fact demonstrated in Fig.~\ref{fig:resx0} is that the two local characteristics quickly converge to constant values independent of initial conditions. The limiting values correspond to local properties of the quasi-stationary distribution, $Q_{\rm st}(x)$. The PDF $Q_{\rm st}(x)$ naturally arises in our experiment and is inherent in all highly unstable systems.

The quasi-stationary distribution is defined as the long-time limit of the normalized position PDF: $Q_{\rm st}(x) ={\displaystyle \lim_{t\to \infty} }P(x,t)/S(t)$, where $P(x,t)$ satisfies the Fokker-Planck equation for the discussed model, and $S(t)=\int_{-\infty}^{+\infty}dx' P(x',t)$ is the survival probability~\cite{bookRedner}, which gives the probability that the trajectory has not diverged by the time $t$. Due to the high instability of the potential, the number of non-diverging trajectories decreases exponentially, $S(t)\sim {\rm e}^{-\lambda_0 t}$, $P(x,t)\sim Q_{\rm st}(x) {\rm e}^{-\lambda_0 t}$, and the normalized PDF $Q(x,t)=P(x,t)/S(t)$ converges exponentially fast to $Q_{\rm st}(x)$.  Introducing the normalized PDF $Q(x,t)$ into the Fokker-Planck equation corresponding to the Langevin equation~\eqref{eq:LE}, yields that $Q_{\rm st}(x)$ is given by the normalized eigenvector for the smallest $\lambda_0$ of the Fokker-Planck operator in question~\cite{OrnigottiPRE18}:  
\begin{equation}
\label{eq:QstDiff}
\frac{1}{\gamma}
 \left[ k_BT \partial_{xx}^{2} + \partial_{x} V'(x) \right] Q_{\rm st}(x) = -  \lambda_0 Q_{\rm st}(x). 
\end{equation} 
Eq.~\eqref{eq:QstDiff} is proven in the mathematical literature on quasi-stationary distributions \cite{bookQSD} and explained on physical grounds in SM~\cite{SI}. Fig.~\ref{fig:PT} demonstrates fast convergence of the measured PDF in the highly unstable cubic potential, $V(x)=\mu x^{3}/3$, towards $Q_{\rm st}(x)$ calculated from Eq.~(\ref{eq:QstDiff}) (with natural boundary conditions). The histogram of measured positions is well approximated by $Q_{\rm st}(x)$ already for short times, $t>1$ s.


Figs.~\ref{fig:PT} and~\ref{fig:resx0} show that the quasi-stationary limit of $x_{\rm max}(t)$, $x_{\rm max} \equiv {\displaystyle \lim_{t\to\infty} } x_{\rm max}(t)$, is shifted right from the origin against the direction of the acting force. This reflects the tendency of {\em long non-diverging trajectories} (the trajectories that have not diverged by the long time $t$, $t\to \infty$) to diffuse right from $x=0$ for the most part of the time. Such long-surviving trajectories tend to avoid negative values of $x$, because an excursion to negative $x$, where the strong force may cause their divergence, would  almost surely  be fatal for them. Thus the long-surviving trajectories are most probably found on the plateau region slightly right from $x=0$.

In order to distinguish this quasi-stationary atypical shift in experimental data, the quasi-stationary value of $\sigma(t)$, $\sigma \equiv {\displaystyle \lim_{t\to\infty} }\sigma(t)$, should be reasonably small compared to  $x_{\rm max}$. Large $\sigma$ corresponds to a broad distribution around the maximum, where it may be hard to measure $x_{\rm max}$ with a sufficient precision. The experimental results demonstrate saturation of the ``shift-to-noise'' ratio  $x_{\rm max}/\sigma = (0.5\pm0.2)$. Solution of Eq.~(\ref{eq:QstDiff}) predicts that $x_{\rm max}/\sigma \approx 0.67$. These values are sufficiently large to clearly observe the atypical shift of $x_{\rm max}$ from the origin. The atypical shift is a robust effect which can be observed even without cooling and for a strong nonlinearity.

\section{Rate of divergence}

The magnitude of the eigenvalue $\lambda_0$ gives a quantitative measure of stability of the studied system. It is equal to the slowest decay rate in the given potential \cite{RiskenBook}, or in our case, the rate at which the trajectories diverge. Evaluating Eq.~(\ref{eq:QstDiff}) at $x=x_{\rm max}$ and identifying the quasi-stationary inverse local curvature $\sigma^2$ according to its definition~\eqref{eq:sigma}, we obtain the relation  
\begin{equation} 
\label{eq:lambda0}
 \lambda_0  = \frac{1}{\gamma} \left[ \frac{k_B T}{\sigma^{2}} -  V''(x_{\rm max}) \right], 
\end{equation} 
which allows to determine $\lambda_0$ directly from stationary local quantities $x_{\rm max}$ and $\sigma$. 
Using $\mu = (4.07 \pm 0.03)\ k_B T$ $\mu \mathrm{m}^{-3}$ estimated from the experimental data, we calculate numerically $Q_{\mathrm{st}}(x)$,  $x_{\rm max}$, and $\sigma$ from Eq.~(\ref{eq:QstDiff}), which yields $\lambda_0 = (0.44 \pm 0.01)$ ${\rm s}^{-1}$. The reciprocal magnitude $1/\lambda_0$ of the eigenvalue is the characteristic (longest) decay time. For our experiment we get $1/\lambda_0 =( 2.27 \pm 0.05)$~s. The decay time describes the asymptotic exponential decrease of the average number of nondivergent trajectories. Its magnitude $\approx 2.3$~s ensures that on average there are still enough trajectories around the time $1.5$~s, where the quasi-stationary state emerges, see Figs.~\ref{fig:PT} and~\ref{fig:resx0} to compare the time scales.

\section{Quasi-stationary equipartition theorem}

The counterintuitive shift of $x_{\rm max}$ from zero suggests an interesting quasi-stationary energetics of the unstable system. As a first step towards the development of such a theory, we have derived a generalized equipartition theorem for the quasi-stationary state. The basic idea behind the theorem is to keep only non-diverging trajectories by a proper post-selection process (other trajectories are discarded).  
For odd unstable potentials $V(x)\sim x^{n}$, $n=3,5,\ldots$, the result reduces to the simple formula for the mean potential energy of the Brownian particle, 
\begin{equation}
\label{eq:equipartition}
\left< V(x) \right>_{{\rm st}, +} = \frac{k_BT}{n} + \frac{ \lambda_0 \gamma}{2n} \left<x^{2} \right>_{{\rm st}, +}.
\end{equation} 
Here, the averages are taken over a sufficiently stable section of the system based on the positive half line $x>0$. That is, over the conditional PDF $Q_{{\rm st}, +}(x)= \theta(x) Q_{{\rm st}}(x)/\int_{0}^{\infty} dx' Q_{{\rm st}}(x')$, which describes the quasi-stationary statistics of the stable region, where the most-likely particle position is located. The result~\eqref{eq:equipartition} is obtained directly from the quasi-stationary Fokker-Planck equation~\eqref{eq:QstDiff} after multiplication of the equation by $x^{2}$ and integration over $x\in (0,+\infty)$. Its general form is derived in SM~\cite{SI}. 

The average potential energy $\left< V(x) \right>_{{\rm st}, +}$ is always higher than that obtained from the corresponding equilibrium equipartition theorem, $\left< V(x) \right>_{\rm eq}=k_BT/n$, for the Gibbs state with the same support: $Q_{\rm eq}(x)=\theta(x) \exp\!\left[-V(x)/k_BT \right] /Z $, where an infinite potential barrier restricts the particle to $x>0$ to protect it against divergence. 
The excess energy in the quasi-stationary state is given by the second term on the right-hand side of Eq.~(\ref{eq:equipartition}). This term is always positive.   
Its experimentally measured value, $(\lambda_0 \gamma/6) \langle x^{2} \rangle_{{\rm st}, +} = (0.05 \pm 0.02)k_B T $, agrees with the theoretical prediction $(0.051 \pm 0.001)k_B T$ computed using Eq.~(\ref{eq:QstDiff}) for the measured $\mu$.

In the conditional ensemble we discard all diverging trajectories with low potential energies located at $x<0$. The excess energy arises due to the heat accepted from the surroundings by non-diverging trajectories. The heat ${\mathcal Q}(t)$, accepted during $(0,t)$, is ${\mathcal Q}(t)=V(x(t))-V(x(0))$, because $V(x)$ is time-independent~\cite{Seifert2012}. 
Remarkably, the quasi-static conditional strategy can perform better in harnessing potential energy compared to the equilibrium one. This opens possibilities for further thermodynamic investigation of work and heat extractable from quasi-stationary states and calls for extension of our method to time-dependent potentials $V(x,t)$, where ${\mathcal Q}(t)$ will no longer be the simple difference of potential energies~\cite{Seifert2012}.

\section{Summary and perspectives}

Our experimental tests successfully verified (i) utility of the approach based on the most-likely motion of the unstable process, (ii) fast appearance of the quasi-stationary distribution for room-temperature overdamped dynamics and (iii) the generalization of the equipartition theorem~(\ref{eq:equipartition}) for a regular part of the quasi-stationary PDF. All experimental results are in good agreement with the theory even for a small number of trajectories. 
We have shown that the naturally arising quasi-stationary state has a higher energetic content than the equilibrium one. After further themodynamic investigation, this finding may stimulate development of new approaches to exploit this advantage and design new thermal machines based on unstable dynamics. From a general perspective, our results suggest a high potential of unstable systems for future applications in microscopic machines. The next experimental challenge is an unstable underdamped regime~\cite{FonsecaPhysRevLett2016, Ricci2017, Rondin2017}, where novel phenomena connected to the most-likely dynamics are expected owing to the fact that the underdamped-overdamped correspondence is frequently broken even for stable dynamics~\cite{MartinezPRL2015,Bodrova2016, ArnoldPRE2018}. Advantageously, the approach can be directly translated to almost unitary unstable dynamics in the quantum regime, where the experiments are currently entering \cite{JainPhysRevLett2016, HoangPhysRevLett2016, Rahman2017}.

\section*{Acknowledgements} 
Authors acknowledge support from the Czech Science Foundation:  M.{\v S}., L.O., O.B., P.J., P.Z., R.F.\ were supported by the project GB14-36681G; A.R. and V.H. were supported by the project 17-06716S. 
 V.H.\ is grateful for the support of Alexander von Humboldt Foundation. L.O.\  is supported by the Palacky University (IGA-PrF-2017-008). L.O.\ and R.F.\ have received national funding from the MEYS under grant agreement No.\ 731473 within QUANTERA ERA-NET cofund in quantum technologies implemented within the European Union’s Horizon 2020 Programme (project TheBlinQC). The research infrastructure was supported by MEYS of the Czech Republic, the Czech Academy of Sciences, and European Commission (LO1212, RVO:68081731, and CZ.1.05/2.1.00/01.0017).

\bibliography{references_experimentII}

\begin{thebibliography}{58}%
\makeatletter
\providecommand \@ifxundefined [1]{%
 \@ifx{#1\undefined}
}%
\providecommand \@ifnum [1]{%
 \ifnum #1\expandafter \@firstoftwo
 \else \expandafter \@secondoftwo
 \fi
}%
\providecommand \@ifx [1]{%
 \ifx #1\expandafter \@firstoftwo
 \else \expandafter \@secondoftwo
 \fi
}%
\providecommand \natexlab [1]{#1}%
\providecommand \enquote  [1]{``#1''}%
\providecommand \bibnamefont  [1]{#1}%
\providecommand \bibfnamefont [1]{#1}%
\providecommand \citenamefont [1]{#1}%
\providecommand \href@noop [0]{\@secondoftwo}%
\providecommand \href [0]{\begingroup \@sanitize@url \@href}%
\providecommand \@href[1]{\@@startlink{#1}\@@href}%
\providecommand \@@href[1]{\endgroup#1\@@endlink}%
\providecommand \@sanitize@url [0]{\catcode `\\12\catcode `\$12\catcode
  `\&12\catcode `\#12\catcode `\^12\catcode `\_12\catcode `\%12\relax}%
\providecommand \@@startlink[1]{}%
\providecommand \@@endlink[0]{}%
\providecommand \url  [0]{\begingroup\@sanitize@url \@url }%
\providecommand \@url [1]{\endgroup\@href {#1}{\urlprefix }}%
\providecommand \urlprefix  [0]{URL }%
\providecommand \Eprint [0]{\href }%
\providecommand \doibase [0]{http://dx.doi.org/}%
\providecommand \selectlanguage [0]{\@gobble}%
\providecommand \bibinfo  [0]{\@secondoftwo}%
\providecommand \bibfield  [0]{\@secondoftwo}%
\providecommand \translation [1]{[#1]}%
\providecommand \BibitemOpen [0]{}%
\providecommand \bibitemStop [0]{}%
\providecommand \bibitemNoStop [0]{.\EOS\space}%
\providecommand \EOS [0]{\spacefactor3000\relax}%
\providecommand \BibitemShut  [1]{\csname bibitem#1\endcsname}%
\let\auto@bib@innerbib\@empty
\bibitem [{\citenamefont {Vale}\ and\ \citenamefont {Milligan}(2000)}]{Vale88}%
  \BibitemOpen
  \bibfield  {author} {\bibinfo {author} {\bibfnamefont {R.~D.}\ \bibnamefont
  {Vale}}\ and\ \bibinfo {author} {\bibfnamefont {R.~A.}\ \bibnamefont
  {Milligan}},\ }\href {\doibase 10.1126/science.288.5463.88} {\bibfield
  {journal} {\bibinfo  {journal} {Science}\ }\textbf {\bibinfo {volume}
  {288}},\ \bibinfo {pages} {88} (\bibinfo {year} {2000})}\BibitemShut
  {NoStop}%
\bibitem [{\citenamefont {Schliwa}\ and\ \citenamefont
  {Woehlke}(2003)}]{Schliwa2003}%
  \BibitemOpen
  \bibfield  {author} {\bibinfo {author} {\bibfnamefont {M.}~\bibnamefont
  {Schliwa}}\ and\ \bibinfo {author} {\bibfnamefont {G.}~\bibnamefont
  {Woehlke}},\ }\href {\doibase 10.1038/nature01601} {\bibfield  {journal}
  {\bibinfo  {journal} {Nature}\ }\textbf {\bibinfo {volume} {422}},\ \bibinfo
  {pages} {759} (\bibinfo {year} {2003})}\BibitemShut {NoStop}%
\bibitem [{\citenamefont {Astumian}(2016)}]{AstumianFarDis2016}%
  \BibitemOpen
  \bibfield  {author} {\bibinfo {author} {\bibfnamefont {R.~D.}\ \bibnamefont
  {Astumian}},\ }\href {\doibase 10.1039/C6FD00140H} {\bibfield  {journal}
  {\bibinfo  {journal} {Faraday Discuss.}\ }\textbf {\bibinfo {volume} {195}},\
  \bibinfo {pages} {583} (\bibinfo {year} {2016})}\BibitemShut {NoStop}%
\bibitem [{\citenamefont {Erbas-Cakmak}\ \emph {et~al.}(2015)\citenamefont
  {Erbas-Cakmak}, \citenamefont {Leigh}, \citenamefont {McTernan},\ and\
  \citenamefont {Nussbaumer}}]{ErbasCakmak2015}%
  \BibitemOpen
  \bibfield  {author} {\bibinfo {author} {\bibfnamefont {S.}~\bibnamefont
  {Erbas-Cakmak}}, \bibinfo {author} {\bibfnamefont {D.~A.}\ \bibnamefont
  {Leigh}}, \bibinfo {author} {\bibfnamefont {C.~T.}\ \bibnamefont {McTernan}},
  \ and\ \bibinfo {author} {\bibfnamefont {A.~L.}\ \bibnamefont {Nussbaumer}},\
  }\href {\doibase 10.1021/acs.chemrev.5b00146} {\bibfield  {journal} {\bibinfo
   {journal} {Chem. Rev.}\ }\textbf {\bibinfo {volume} {115}},\ \bibinfo
  {pages} {10081} (\bibinfo {year} {2015})}\BibitemShut {NoStop}%
\bibitem [{\citenamefont {Filip}\ and\ \citenamefont {Zem{\'
  a}nek}(2016)}]{FilipJOpt16}%
  \BibitemOpen
  \bibfield  {author} {\bibinfo {author} {\bibfnamefont {R.}~\bibnamefont
  {Filip}}\ and\ \bibinfo {author} {\bibfnamefont {P.}~\bibnamefont {Zem{\'
  a}nek}},\ }\href {http://stacks.iop.org/2040-8986/18/i=6/a=065401} {\bibfield
   {journal} {\bibinfo  {journal} {J. Opt.}\ }\textbf {\bibinfo {volume}
  {18}},\ \bibinfo {pages} {065401} (\bibinfo {year} {2016})}\BibitemShut
  {NoStop}%
\bibitem [{\citenamefont {Ornigotti}\ \emph {et~al.}(2018)\citenamefont
  {Ornigotti}, \citenamefont {Ryabov}, \citenamefont {Holubec},\ and\
  \citenamefont {Filip}}]{OrnigottiPRE18}%
  \BibitemOpen
  \bibfield  {author} {\bibinfo {author} {\bibfnamefont {L.}~\bibnamefont
  {Ornigotti}}, \bibinfo {author} {\bibfnamefont {A.}~\bibnamefont {Ryabov}},
  \bibinfo {author} {\bibfnamefont {V.}~\bibnamefont {Holubec}}, \ and\
  \bibinfo {author} {\bibfnamefont {R.}~\bibnamefont {Filip}},\ }\href
  {\doibase 10.1103/PhysRevE.97.032127} {\bibfield  {journal} {\bibinfo
  {journal} {Phys. Rev. E}\ }\textbf {\bibinfo {volume} {97}},\ \bibinfo
  {pages} {032127} (\bibinfo {year} {2018})}\BibitemShut {NoStop}%
\bibitem [{\citenamefont {Sigeti}\ and\ \citenamefont
  {Horsthemke}(1989)}]{Sigeti1989}%
  \BibitemOpen
  \bibfield  {author} {\bibinfo {author} {\bibfnamefont {D.}~\bibnamefont
  {Sigeti}}\ and\ \bibinfo {author} {\bibfnamefont {W.}~\bibnamefont
  {Horsthemke}},\ }\href {\doibase 10.1007/BF01044713} {\bibfield  {journal}
  {\bibinfo  {journal} {J. Stat. Phys.}\ }\textbf {\bibinfo {volume} {54}},\
  \bibinfo {pages} {1217} (\bibinfo {year} {1989})}\BibitemShut {NoStop}%
\bibitem [{\citenamefont {Reimann}\ and\ \citenamefont {{{Van den
  Broeck}}}(1994)}]{ReimannBroeck94}%
  \BibitemOpen
  \bibfield  {author} {\bibinfo {author} {\bibfnamefont {P.}~\bibnamefont
  {Reimann}}\ and\ \bibinfo {author} {\bibfnamefont {C.}~\bibnamefont {{{Van
  den Broeck}}}},\ }\href {\doibase 10.1016/0167-2789(94)00095-6} {\bibfield
  {journal} {\bibinfo  {journal} {Physica D: Nonlinear Phenomena}\ }\textbf
  {\bibinfo {volume} {75}},\ \bibinfo {pages} {509} (\bibinfo {year}
  {1994})}\BibitemShut {NoStop}%
\bibitem [{\citenamefont {Colet}\ \emph {et~al.}(1989)\citenamefont {Colet},
  \citenamefont {San~Miguel}, \citenamefont {Casademunt},\ and\ \citenamefont
  {Sancho}}]{SanchoPRA1989}%
  \BibitemOpen
  \bibfield  {author} {\bibinfo {author} {\bibfnamefont {P.}~\bibnamefont
  {Colet}}, \bibinfo {author} {\bibfnamefont {M.}~\bibnamefont {San~Miguel}},
  \bibinfo {author} {\bibfnamefont {J.}~\bibnamefont {Casademunt}}, \ and\
  \bibinfo {author} {\bibfnamefont {J.~M.}\ \bibnamefont {Sancho}},\ }\href
  {\doibase 10.1103/PhysRevA.39.149} {\bibfield  {journal} {\bibinfo  {journal}
  {Phys. Rev. A}\ }\textbf {\bibinfo {volume} {39}},\ \bibinfo {pages} {149}
  (\bibinfo {year} {1989})}\BibitemShut {NoStop}%
\bibitem [{\citenamefont {Ram{\' i}rez-Piscina}\ and\ \citenamefont
  {Sancho}(1991)}]{SanchoPRA1991}%
  \BibitemOpen
  \bibfield  {author} {\bibinfo {author} {\bibfnamefont {L.}~\bibnamefont
  {Ram{\' i}rez-Piscina}}\ and\ \bibinfo {author} {\bibfnamefont {J.~M.}\
  \bibnamefont {Sancho}},\ }\href {\doibase 10.1103/PhysRevA.43.663} {\bibfield
   {journal} {\bibinfo  {journal} {Phys. Rev. A}\ }\textbf {\bibinfo {volume}
  {43}},\ \bibinfo {pages} {663} (\bibinfo {year} {1991})}\BibitemShut
  {NoStop}%
\bibitem [{\citenamefont {Arecchi}\ and\ \citenamefont
  {Degiorgio}(1971)}]{ArecchiPRA1971}%
  \BibitemOpen
  \bibfield  {author} {\bibinfo {author} {\bibfnamefont {F.~T.}\ \bibnamefont
  {Arecchi}}\ and\ \bibinfo {author} {\bibfnamefont {V.}~\bibnamefont
  {Degiorgio}},\ }\href {\doibase 10.1103/PhysRevA.3.1108} {\bibfield
  {journal} {\bibinfo  {journal} {Phys. Rev. A}\ }\textbf {\bibinfo {volume}
  {3}},\ \bibinfo {pages} {1108} (\bibinfo {year} {1971})}\BibitemShut
  {NoStop}%
\bibitem [{\citenamefont {Lindner}\ \emph {et~al.}(2003)\citenamefont
  {Lindner}, \citenamefont {Longtin},\ and\ \citenamefont
  {Bulsara}}]{Lindner2003}%
  \BibitemOpen
  \bibfield  {author} {\bibinfo {author} {\bibfnamefont {B.}~\bibnamefont
  {Lindner}}, \bibinfo {author} {\bibfnamefont {A.}~\bibnamefont {Longtin}}, \
  and\ \bibinfo {author} {\bibfnamefont {A.}~\bibnamefont {Bulsara}},\ }\href
  {\doibase 10.1162/08997660360675035} {\bibfield  {journal} {\bibinfo
  {journal} {Neural Comput.}\ }\textbf {\bibinfo {volume} {15}},\ \bibinfo
  {pages} {1761} (\bibinfo {year} {2003})}\BibitemShut {NoStop}%
\bibitem [{\citenamefont {Brunel}\ and\ \citenamefont
  {Latham}(2003)}]{Brunel2003}%
  \BibitemOpen
  \bibfield  {author} {\bibinfo {author} {\bibfnamefont {N.}~\bibnamefont
  {Brunel}}\ and\ \bibinfo {author} {\bibfnamefont {P.~E.}\ \bibnamefont
  {Latham}},\ }\href {\doibase 10.1162/089976603322362365} {\bibfield
  {journal} {\bibinfo  {journal} {Neural Comput.}\ }\textbf {\bibinfo {volume}
  {15}},\ \bibinfo {pages} {2281} (\bibinfo {year} {2003})}\BibitemShut
  {NoStop}%
\bibitem [{\citenamefont {Reimann}\ \emph {et~al.}(2001)\citenamefont
  {Reimann}, \citenamefont {Van~den Broeck}, \citenamefont {Linke},
  \citenamefont {H\"anggi}, \citenamefont {Rubi},\ and\ \citenamefont
  {P\'erez-Madrid}}]{ReimannPRL01}%
  \BibitemOpen
  \bibfield  {author} {\bibinfo {author} {\bibfnamefont {P.}~\bibnamefont
  {Reimann}}, \bibinfo {author} {\bibfnamefont {C.}~\bibnamefont {Van~den
  Broeck}}, \bibinfo {author} {\bibfnamefont {H.}~\bibnamefont {Linke}},
  \bibinfo {author} {\bibfnamefont {P.}~\bibnamefont {H\"anggi}}, \bibinfo
  {author} {\bibfnamefont {J.~M.}\ \bibnamefont {Rubi}}, \ and\ \bibinfo
  {author} {\bibfnamefont {A.}~\bibnamefont {P\'erez-Madrid}},\ }\href
  {\doibase 10.1103/PhysRevLett.87.010602} {\bibfield  {journal} {\bibinfo
  {journal} {Phys. Rev. Lett.}\ }\textbf {\bibinfo {volume} {87}},\ \bibinfo
  {pages} {010602} (\bibinfo {year} {2001})}\BibitemShut {NoStop}%
\bibitem [{\citenamefont {Reimann}\ \emph {et~al.}(2002)\citenamefont
  {Reimann}, \citenamefont {Van~den Broeck}, \citenamefont {Linke},
  \citenamefont {H\"anggi}, \citenamefont {Rubi},\ and\ \citenamefont
  {P\'erez-Madrid}}]{ReimannPRE02}%
  \BibitemOpen
  \bibfield  {author} {\bibinfo {author} {\bibfnamefont {P.}~\bibnamefont
  {Reimann}}, \bibinfo {author} {\bibfnamefont {C.}~\bibnamefont {Van~den
  Broeck}}, \bibinfo {author} {\bibfnamefont {H.}~\bibnamefont {Linke}},
  \bibinfo {author} {\bibfnamefont {P.}~\bibnamefont {H\"anggi}}, \bibinfo
  {author} {\bibfnamefont {J.~M.}\ \bibnamefont {Rubi}}, \ and\ \bibinfo
  {author} {\bibfnamefont {A.}~\bibnamefont {P\'erez-Madrid}},\ }\href
  {\doibase 10.1103/PhysRevE.65.031104} {\bibfield  {journal} {\bibinfo
  {journal} {Phys. Rev. E}\ }\textbf {\bibinfo {volume} {65}},\ \bibinfo
  {pages} {031104} (\bibinfo {year} {2002})}\BibitemShut {NoStop}%
\bibitem [{\citenamefont {Gu\'erin}\ and\ \citenamefont {Dean}(2017)}]{Dean1}%
  \BibitemOpen
  \bibfield  {author} {\bibinfo {author} {\bibfnamefont {T.}~\bibnamefont
  {Gu\'erin}}\ and\ \bibinfo {author} {\bibfnamefont {D.~S.}\ \bibnamefont
  {Dean}},\ }\href {\doibase 10.1103/PhysRevE.95.012109} {\bibfield  {journal}
  {\bibinfo  {journal} {Phys. Rev. E}\ }\textbf {\bibinfo {volume} {95}},\
  \bibinfo {pages} {012109} (\bibinfo {year} {2017})}\BibitemShut {NoStop}%
\bibitem [{\citenamefont {Arzola}\ \emph {et~al.}(2017)\citenamefont {Arzola},
  \citenamefont {Villasante-Barahona}, \citenamefont {Volke-Sep\'ulveda},
  \citenamefont {J\'akl},\ and\ \citenamefont {Zem\'anek}}]{PavelPRL2017}%
  \BibitemOpen
  \bibfield  {author} {\bibinfo {author} {\bibfnamefont {A.~V.}\ \bibnamefont
  {Arzola}}, \bibinfo {author} {\bibfnamefont {M.}~\bibnamefont
  {Villasante-Barahona}}, \bibinfo {author} {\bibfnamefont {K.}~\bibnamefont
  {Volke-Sep\'ulveda}}, \bibinfo {author} {\bibfnamefont {P.}~\bibnamefont
  {J\'akl}}, \ and\ \bibinfo {author} {\bibfnamefont {P.}~\bibnamefont
  {Zem\'anek}},\ }\href {\doibase 10.1103/PhysRevLett.118.138002} {\bibfield
  {journal} {\bibinfo  {journal} {Phys. Rev. Lett.}\ }\textbf {\bibinfo
  {volume} {118}},\ \bibinfo {pages} {138002} (\bibinfo {year}
  {2017})}\BibitemShut {NoStop}%
\bibitem [{\citenamefont {Hirsch}\ \emph {et~al.}(1982)\citenamefont {Hirsch},
  \citenamefont {Huberman},\ and\ \citenamefont {Scalapino}}]{HirschPRA1982}%
  \BibitemOpen
  \bibfield  {author} {\bibinfo {author} {\bibfnamefont {J.~E.}\ \bibnamefont
  {Hirsch}}, \bibinfo {author} {\bibfnamefont {B.~A.}\ \bibnamefont
  {Huberman}}, \ and\ \bibinfo {author} {\bibfnamefont {D.~J.}\ \bibnamefont
  {Scalapino}},\ }\href {\doibase 10.1103/PhysRevA.25.519} {\bibfield
  {journal} {\bibinfo  {journal} {Phys. Rev. A}\ }\textbf {\bibinfo {volume}
  {25}},\ \bibinfo {pages} {519} (\bibinfo {year} {1982})}\BibitemShut
  {NoStop}%
\bibitem [{SI()}]{SI}%
  \BibitemOpen
  \href@noop {} {}\bibinfo {note} {See Supplemental Material submitted as the
  Ancillary file, including Refs.~\cite{LamourouxPLA09, OHaganBiometrika76,
  TolmanBook, Tolman1918, HuangBOOK, SekimotoBOOK}.}\BibitemShut {Stop}%
\bibitem [{\citenamefont {Tretiakov}\ \emph {et~al.}(2003)\citenamefont
  {Tretiakov}, \citenamefont {Gramespacher},\ and\ \citenamefont
  {Matveev}}]{TretiakovPRB2003}%
  \BibitemOpen
  \bibfield  {author} {\bibinfo {author} {\bibfnamefont {O.~A.}\ \bibnamefont
  {Tretiakov}}, \bibinfo {author} {\bibfnamefont {T.}~\bibnamefont
  {Gramespacher}}, \ and\ \bibinfo {author} {\bibfnamefont {K.~A.}\
  \bibnamefont {Matveev}},\ }\href {\doibase 10.1103/PhysRevB.67.073303}
  {\bibfield  {journal} {\bibinfo  {journal} {Phys. Rev. B}\ }\textbf {\bibinfo
  {volume} {67}},\ \bibinfo {pages} {073303} (\bibinfo {year}
  {2003})}\BibitemShut {NoStop}%
\bibitem [{\citenamefont {Spagnolo}\ \emph {et~al.}(2017)\citenamefont
  {Spagnolo}, \citenamefont {Guarcello}, \citenamefont {Magazzù},
  \citenamefont {Carollo}, \citenamefont {Persano~Adorno},\ and\ \citenamefont
  {Valenti}}]{SpagnoloEntropy2017}%
  \BibitemOpen
  \bibfield  {author} {\bibinfo {author} {\bibfnamefont {B.}~\bibnamefont
  {Spagnolo}}, \bibinfo {author} {\bibfnamefont {C.}~\bibnamefont {Guarcello}},
  \bibinfo {author} {\bibfnamefont {L.}~\bibnamefont {Magazzù}}, \bibinfo
  {author} {\bibfnamefont {A.}~\bibnamefont {Carollo}}, \bibinfo {author}
  {\bibfnamefont {D.}~\bibnamefont {Persano~Adorno}}, \ and\ \bibinfo {author}
  {\bibfnamefont {D.}~\bibnamefont {Valenti}},\ }\href {\doibase
  10.3390/e19010020} {\bibfield  {journal} {\bibinfo  {journal} {Entropy}\
  }\textbf {\bibinfo {volume} {19}},\ \bibinfo {pages} {20} (\bibinfo {year}
  {2017})}\BibitemShut {NoStop}%
\bibitem [{\citenamefont {Arecchi}\ \emph {et~al.}(1982)\citenamefont
  {Arecchi}, \citenamefont {Politi},\ and\ \citenamefont
  {Ulivi}}]{Arecchi1982}%
  \BibitemOpen
  \bibfield  {author} {\bibinfo {author} {\bibfnamefont {F.~T.}\ \bibnamefont
  {Arecchi}}, \bibinfo {author} {\bibfnamefont {A.}~\bibnamefont {Politi}}, \
  and\ \bibinfo {author} {\bibfnamefont {L.}~\bibnamefont {Ulivi}},\ }\href
  {\doibase 10.1007/BF02721698} {\bibfield  {journal} {\bibinfo  {journal} {Il
  Nuovo Cimento B (1971-1996)}\ }\textbf {\bibinfo {volume} {71}},\ \bibinfo
  {pages} {119} (\bibinfo {year} {1982})}\BibitemShut {NoStop}%
\bibitem [{\citenamefont {Young}\ and\ \citenamefont
  {Singh}(1985)}]{YoungPRA1985}%
  \BibitemOpen
  \bibfield  {author} {\bibinfo {author} {\bibfnamefont {M.~R.}\ \bibnamefont
  {Young}}\ and\ \bibinfo {author} {\bibfnamefont {S.}~\bibnamefont {Singh}},\
  }\href {\doibase 10.1103/PhysRevA.31.888} {\bibfield  {journal} {\bibinfo
  {journal} {Phys. Rev. A}\ }\textbf {\bibinfo {volume} {31}},\ \bibinfo
  {pages} {888} (\bibinfo {year} {1985})}\BibitemShut {NoStop}%
\bibitem [{\citenamefont {C{\' a}ceres}\ \emph {et~al.}(1995)\citenamefont
  {C{\' a}ceres}, \citenamefont {Budde},\ and\ \citenamefont
  {Sibona}}]{Caceres1995}%
  \BibitemOpen
  \bibfield  {author} {\bibinfo {author} {\bibfnamefont {M.~O.}\ \bibnamefont
  {C{\' a}ceres}}, \bibinfo {author} {\bibfnamefont {C.~E.}\ \bibnamefont
  {Budde}}, \ and\ \bibinfo {author} {\bibfnamefont {G.~J.}\ \bibnamefont
  {Sibona}},\ }\href {http://stacks.iop.org/0305-4470/28/i=14/a=009} {\bibfield
   {journal} {\bibinfo  {journal} {J. Phys. A: Math. Gen.}\ }\textbf {\bibinfo
  {volume} {28}},\ \bibinfo {pages} {3877} (\bibinfo {year}
  {1995})}\BibitemShut {NoStop}%
\bibitem [{\citenamefont {Mantegna}\ and\ \citenamefont
  {Spagnolo}(1996)}]{MantegnaPRL1996}%
  \BibitemOpen
  \bibfield  {author} {\bibinfo {author} {\bibfnamefont {R.~N.}\ \bibnamefont
  {Mantegna}}\ and\ \bibinfo {author} {\bibfnamefont {B.}~\bibnamefont
  {Spagnolo}},\ }\href {\doibase 10.1103/PhysRevLett.76.563} {\bibfield
  {journal} {\bibinfo  {journal} {Phys. Rev. Lett.}\ }\textbf {\bibinfo
  {volume} {76}},\ \bibinfo {pages} {563} (\bibinfo {year} {1996})}\BibitemShut
  {NoStop}%
\bibitem [{\citenamefont {Agudov}(1998)}]{AgudovPRE1998}%
  \BibitemOpen
  \bibfield  {author} {\bibinfo {author} {\bibfnamefont {N.~V.}\ \bibnamefont
  {Agudov}},\ }\href {\doibase 10.1103/PhysRevE.57.2618} {\bibfield  {journal}
  {\bibinfo  {journal} {Phys. Rev. E}\ }\textbf {\bibinfo {volume} {57}},\
  \bibinfo {pages} {2618} (\bibinfo {year} {1998})}\BibitemShut {NoStop}%
\bibitem [{\citenamefont {Fiasconaro}\ \emph {et~al.}(2005)\citenamefont
  {Fiasconaro}, \citenamefont {Spagnolo},\ and\ \citenamefont
  {Boccaletti}}]{FiasconaroPRE2005}%
  \BibitemOpen
  \bibfield  {author} {\bibinfo {author} {\bibfnamefont {A.}~\bibnamefont
  {Fiasconaro}}, \bibinfo {author} {\bibfnamefont {B.}~\bibnamefont
  {Spagnolo}}, \ and\ \bibinfo {author} {\bibfnamefont {S.}~\bibnamefont
  {Boccaletti}},\ }\href {\doibase 10.1103/PhysRevE.72.061110} {\bibfield
  {journal} {\bibinfo  {journal} {Phys. Rev. E}\ }\textbf {\bibinfo {volume}
  {72}},\ \bibinfo {pages} {061110} (\bibinfo {year} {2005})}\BibitemShut
  {NoStop}%
\bibitem [{\citenamefont {C{\'a}ceres}(2008)}]{CaceresJSP2008}%
  \BibitemOpen
  \bibfield  {author} {\bibinfo {author} {\bibfnamefont {M.~O.}\ \bibnamefont
  {C{\'a}ceres}},\ }\href {\doibase 10.1007/s10955-008-9554-7} {\bibfield
  {journal} {\bibinfo  {journal} {J. Stat. Phys.}\ }\textbf {\bibinfo {volume}
  {132}},\ \bibinfo {pages} {487} (\bibinfo {year} {2008})}\BibitemShut
  {NoStop}%
\bibitem [{\citenamefont {Ryabov}\ \emph {et~al.}(2016)\citenamefont {Ryabov},
  \citenamefont {Zem\'anek},\ and\ \citenamefont {Filip}}]{RyabovPRE16}%
  \BibitemOpen
  \bibfield  {author} {\bibinfo {author} {\bibfnamefont {A.}~\bibnamefont
  {Ryabov}}, \bibinfo {author} {\bibfnamefont {P.}~\bibnamefont {Zem\'anek}}, \
  and\ \bibinfo {author} {\bibfnamefont {R.}~\bibnamefont {Filip}},\ }\href
  {\doibase 10.1103/PhysRevE.94.042108} {\bibfield  {journal} {\bibinfo
  {journal} {Phys. Rev. E}\ }\textbf {\bibinfo {volume} {94}},\ \bibinfo
  {pages} {042108} (\bibinfo {year} {2016})}\BibitemShut {NoStop}%
\bibitem [{\citenamefont {H{\" a}nggi}\ \emph {et~al.}(1990)\citenamefont {H{\"
  a}nggi}, \citenamefont {Talkner},\ and\ \citenamefont
  {Borkovec}}]{HanggiRevModPhys1990}%
  \BibitemOpen
  \bibfield  {author} {\bibinfo {author} {\bibfnamefont {P.}~\bibnamefont {H{\"
  a}nggi}}, \bibinfo {author} {\bibfnamefont {P.}~\bibnamefont {Talkner}}, \
  and\ \bibinfo {author} {\bibfnamefont {M.}~\bibnamefont {Borkovec}},\ }\href
  {\doibase 10.1103/RevModPhys.62.251} {\bibfield  {journal} {\bibinfo
  {journal} {Rev. Mod. Phys.}\ }\textbf {\bibinfo {volume} {62}},\ \bibinfo
  {pages} {251} (\bibinfo {year} {1990})}\BibitemShut {NoStop}%
\bibitem [{\citenamefont {{\v S}iler}\ \emph {et~al.}(2017)\citenamefont {{\v
  S}iler}, \citenamefont {J{\' a}kl}, \citenamefont {Brzobohatý},
  \citenamefont {Ryabov}, \citenamefont {Filip},\ and\ \citenamefont {Zem{\'
  a}nek}}]{SilerSR17}%
  \BibitemOpen
  \bibfield  {author} {\bibinfo {author} {\bibfnamefont {M.}~\bibnamefont {{\v
  S}iler}}, \bibinfo {author} {\bibfnamefont {P.}~\bibnamefont {J{\' a}kl}},
  \bibinfo {author} {\bibfnamefont {O.}~\bibnamefont {Brzobohatý}}, \bibinfo
  {author} {\bibfnamefont {A.}~\bibnamefont {Ryabov}}, \bibinfo {author}
  {\bibfnamefont {R.}~\bibnamefont {Filip}}, \ and\ \bibinfo {author}
  {\bibfnamefont {P.}~\bibnamefont {Zem{\' a}nek}},\ }\href {\doibase
  10.1038/s41598-017-01848-4} {\bibfield  {journal} {\bibinfo  {journal} {Sci.
  Rep.}\ }\textbf {\bibinfo {volume} {7}},\ \bibinfo {pages} {1697} (\bibinfo
  {year} {2017})}\BibitemShut {NoStop}%
\bibitem [{\citenamefont {Yaglom}(1947)}]{Yaglom1947}%
  \BibitemOpen
  \bibfield  {author} {\bibinfo {author} {\bibfnamefont {A.~M.}\ \bibnamefont
  {Yaglom}},\ }\href@noop {} {\bibfield  {journal} {\bibinfo  {journal} {Dokl.
  Acad. Nauk SSSR (in Russian)}\ }\textbf {\bibinfo {volume} {56}},\ \bibinfo
  {pages} {795} (\bibinfo {year} {1947})}\BibitemShut {NoStop}%
\bibitem [{\citenamefont {Collet}\ \emph {et~al.}(2013)\citenamefont {Collet},
  \citenamefont {Mart{\' i}nez},\ and\ \citenamefont {{San Mart{\'
  i}n}}}]{bookQSD}%
  \BibitemOpen
  \bibfield  {author} {\bibinfo {author} {\bibfnamefont {P.}~\bibnamefont
  {Collet}}, \bibinfo {author} {\bibfnamefont {S.}~\bibnamefont {Mart{\'
  i}nez}}, \ and\ \bibinfo {author} {\bibfnamefont {J.}~\bibnamefont {{San
  Mart{\' i}n}}},\ }\href@noop {} {\emph {\bibinfo {title} {Quasi-Stationary
  Distributions: Markov Chains, Diffusions and Dynamical Systems}}}\ (\bibinfo
  {publisher} {Springer-Verlag Berlin Heidelberg},\ \bibinfo {year}
  {2013})\BibitemShut {NoStop}%
\bibitem [{\citenamefont {Pollett}(2015)}]{PolletURL}%
  \BibitemOpen
  \bibfield  {author} {\bibinfo {author} {\bibfnamefont {P.~K.}\ \bibnamefont
  {Pollett}},\ }\href
  {https://people.smp.uq.edu.au/PhilipPollett/papers/qsds/qsds.pdf} {\enquote
  {\bibinfo {title} {Quasi-stationary distributions: A bibliography},}\ }
  (\bibinfo {year} {2015})\BibitemShut {NoStop}%
\bibitem [{\citenamefont {N{\aa}sell}(1995)}]{Nasell1995}%
  \BibitemOpen
  \bibfield  {author} {\bibinfo {author} {\bibfnamefont {I.}~\bibnamefont
  {N{\aa}sell}},\ }\href {\doibase 10.2307/1428186} {\bibfield  {journal}
  {\bibinfo  {journal} {Adv. Appl. Probab.}\ }\textbf {\bibinfo {volume}
  {28}},\ \bibinfo {pages} {895} (\bibinfo {year} {1995})}\BibitemShut
  {NoStop}%
\bibitem [{\citenamefont {Hastings}(2004)}]{Hastings2004}%
  \BibitemOpen
  \bibfield  {author} {\bibinfo {author} {\bibfnamefont {A.}~\bibnamefont
  {Hastings}},\ }\href {\doibase 10.1016/j.tree.2003.09.007} {\bibfield
  {journal} {\bibinfo  {journal} {Trends Ecol. Evol.}\ }\textbf {\bibinfo
  {volume} {19}},\ \bibinfo {pages} {39} (\bibinfo {year} {2004})}\BibitemShut
  {NoStop}%
\bibitem [{\citenamefont {Steinsaltz}\ and\ \citenamefont
  {Evans}(2004)}]{Steinsaltz2004}%
  \BibitemOpen
  \bibfield  {author} {\bibinfo {author} {\bibfnamefont {D.}~\bibnamefont
  {Steinsaltz}}\ and\ \bibinfo {author} {\bibfnamefont {S.~N.}\ \bibnamefont
  {Evans}},\ }\href {\doibase 10.1016/j.tpb.2003.10.007} {\bibfield  {journal}
  {\bibinfo  {journal} {Theor. Pop. Biol.}\ }\textbf {\bibinfo {volume} {65}},\
  \bibinfo {pages} {319} (\bibinfo {year} {2004})}\BibitemShut {NoStop}%
\bibitem [{\citenamefont {Ryabov}\ and\ \citenamefont
  {Chvosta}(2014)}]{TracerPRE2014}%
  \BibitemOpen
  \bibfield  {author} {\bibinfo {author} {\bibfnamefont {A.}~\bibnamefont
  {Ryabov}}\ and\ \bibinfo {author} {\bibfnamefont {P.}~\bibnamefont
  {Chvosta}},\ }\href {\doibase 10.1103/PhysRevE.89.022132} {\bibfield
  {journal} {\bibinfo  {journal} {Phys. Rev. E}\ }\textbf {\bibinfo {volume}
  {89}},\ \bibinfo {pages} {022132} (\bibinfo {year} {2014})}\BibitemShut
  {NoStop}%
\bibitem [{\citenamefont {Fonseca}\ \emph {et~al.}(2016)\citenamefont
  {Fonseca}, \citenamefont {Aranas}, \citenamefont {Millen}, \citenamefont
  {Monteiro},\ and\ \citenamefont {Barker}}]{FonsecaPhysRevLett2016}%
  \BibitemOpen
  \bibfield  {author} {\bibinfo {author} {\bibfnamefont {P.~Z.~G.}\
  \bibnamefont {Fonseca}}, \bibinfo {author} {\bibfnamefont {E.~B.}\
  \bibnamefont {Aranas}}, \bibinfo {author} {\bibfnamefont {J.}~\bibnamefont
  {Millen}}, \bibinfo {author} {\bibfnamefont {T.~S.}\ \bibnamefont
  {Monteiro}}, \ and\ \bibinfo {author} {\bibfnamefont {P.~F.}\ \bibnamefont
  {Barker}},\ }\href {\doibase 10.1103/PhysRevLett.117.173602} {\bibfield
  {journal} {\bibinfo  {journal} {Phys. Rev. Lett.}\ }\textbf {\bibinfo
  {volume} {117}},\ \bibinfo {pages} {173602} (\bibinfo {year}
  {2016})}\BibitemShut {NoStop}%
\bibitem [{\citenamefont {Ricci}\ \emph {et~al.}(2017)\citenamefont {Ricci},
  \citenamefont {Rica}, \citenamefont {Spasenovi{\' c}}, \citenamefont
  {Gieseler}, \citenamefont {Rondin}, \citenamefont {Novotny},\ and\
  \citenamefont {Quidant}}]{Ricci2017}%
  \BibitemOpen
  \bibfield  {author} {\bibinfo {author} {\bibfnamefont {F.}~\bibnamefont
  {Ricci}}, \bibinfo {author} {\bibfnamefont {R.}~\bibnamefont {Rica}},
  \bibinfo {author} {\bibfnamefont {M.}~\bibnamefont {Spasenovi{\' c}}},
  \bibinfo {author} {\bibfnamefont {J.}~\bibnamefont {Gieseler}}, \bibinfo
  {author} {\bibfnamefont {L.}~\bibnamefont {Rondin}}, \bibinfo {author}
  {\bibfnamefont {L.}~\bibnamefont {Novotny}}, \ and\ \bibinfo {author}
  {\bibfnamefont {R.}~\bibnamefont {Quidant}},\ }\href {\doibase
  10.1038/ncomms15141} {\bibfield  {journal} {\bibinfo  {journal} {Nat.
  Commun.}\ }\textbf {\bibinfo {volume} {8}},\ \bibinfo {pages} {15141}
  (\bibinfo {year} {2017})}\BibitemShut {NoStop}%
\bibitem [{\citenamefont {Rondin}\ \emph {et~al.}(2017)\citenamefont {Rondin},
  \citenamefont {Gieseler}, \citenamefont {Ricci}, \citenamefont {Quidant},
  \citenamefont {Dellago},\ and\ \citenamefont {Novotny}}]{Rondin2017}%
  \BibitemOpen
  \bibfield  {author} {\bibinfo {author} {\bibfnamefont {L.}~\bibnamefont
  {Rondin}}, \bibinfo {author} {\bibfnamefont {J.}~\bibnamefont {Gieseler}},
  \bibinfo {author} {\bibfnamefont {F.}~\bibnamefont {Ricci}}, \bibinfo
  {author} {\bibfnamefont {R.}~\bibnamefont {Quidant}}, \bibinfo {author}
  {\bibfnamefont {C.}~\bibnamefont {Dellago}}, \ and\ \bibinfo {author}
  {\bibfnamefont {L.}~\bibnamefont {Novotny}},\ }\href
  {http://dx.doi.org/10.1038/nnano.2017.198} {\bibfield  {journal} {\bibinfo
  {journal} {Nature Nanotechnology}\ }\textbf {\bibinfo {volume} {12}},\
  \bibinfo {pages} {1130} (\bibinfo {year} {2017})}\BibitemShut {NoStop}%
\bibitem [{\citenamefont {Jain}\ \emph {et~al.}(2016)\citenamefont {Jain},
  \citenamefont {Gieseler}, \citenamefont {Moritz}, \citenamefont {Dellago},
  \citenamefont {Quidant},\ and\ \citenamefont
  {Novotny}}]{JainPhysRevLett2016}%
  \BibitemOpen
  \bibfield  {author} {\bibinfo {author} {\bibfnamefont {V.}~\bibnamefont
  {Jain}}, \bibinfo {author} {\bibfnamefont {J.}~\bibnamefont {Gieseler}},
  \bibinfo {author} {\bibfnamefont {C.}~\bibnamefont {Moritz}}, \bibinfo
  {author} {\bibfnamefont {C.}~\bibnamefont {Dellago}}, \bibinfo {author}
  {\bibfnamefont {R.}~\bibnamefont {Quidant}}, \ and\ \bibinfo {author}
  {\bibfnamefont {L.}~\bibnamefont {Novotny}},\ }\href {\doibase
  10.1103/PhysRevLett.116.243601} {\bibfield  {journal} {\bibinfo  {journal}
  {Phys. Rev. Lett.}\ }\textbf {\bibinfo {volume} {116}},\ \bibinfo {pages}
  {243601} (\bibinfo {year} {2016})}\BibitemShut {NoStop}%
\bibitem [{\citenamefont {Hoang}\ \emph {et~al.}(2016)\citenamefont {Hoang},
  \citenamefont {Ma}, \citenamefont {Ahn}, \citenamefont {Bang}, \citenamefont
  {Robicheaux}, \citenamefont {Yin},\ and\ \citenamefont
  {Li}}]{HoangPhysRevLett2016}%
  \BibitemOpen
  \bibfield  {author} {\bibinfo {author} {\bibfnamefont {T.~M.}\ \bibnamefont
  {Hoang}}, \bibinfo {author} {\bibfnamefont {Y.}~\bibnamefont {Ma}}, \bibinfo
  {author} {\bibfnamefont {J.}~\bibnamefont {Ahn}}, \bibinfo {author}
  {\bibfnamefont {J.}~\bibnamefont {Bang}}, \bibinfo {author} {\bibfnamefont
  {F.}~\bibnamefont {Robicheaux}}, \bibinfo {author} {\bibfnamefont {Z.-Q.}\
  \bibnamefont {Yin}}, \ and\ \bibinfo {author} {\bibfnamefont
  {T.}~\bibnamefont {Li}},\ }\href {\doibase 10.1103/PhysRevLett.117.123604}
  {\bibfield  {journal} {\bibinfo  {journal} {Phys. Rev. Lett.}\ }\textbf
  {\bibinfo {volume} {117}},\ \bibinfo {pages} {123604} (\bibinfo {year}
  {2016})}\BibitemShut {NoStop}%
\bibitem [{\citenamefont {Rahman}\ and\ \citenamefont
  {Barker}(2017)}]{Rahman2017}%
  \BibitemOpen
  \bibfield  {author} {\bibinfo {author} {\bibfnamefont {A.~T. M.~A.}\
  \bibnamefont {Rahman}}\ and\ \bibinfo {author} {\bibfnamefont {P.~F.}\
  \bibnamefont {Barker}},\ }\href {\doibase 10.1038/s41566-017-0005-3}
  {\bibfield  {journal} {\bibinfo  {journal} {Nature Photonics}\ }\textbf
  {\bibinfo {volume} {11}},\ \bibinfo {pages} {634} (\bibinfo {year}
  {2017})}\BibitemShut {NoStop}%
\bibitem [{\citenamefont {\v{C}i\v{z}m{\'a}r}\ \emph
  {et~al.}(2011)\citenamefont {\v{C}i\v{z}m{\'a}r}, \citenamefont
  {Brzobohat{\'y}}, \citenamefont {Dholakia},\ and\ \citenamefont
  {Zem{\'a}nek}}]{CizmarLPL11}%
  \BibitemOpen
  \bibfield  {author} {\bibinfo {author} {\bibfnamefont {T.}~\bibnamefont
  {\v{C}i\v{z}m{\'a}r}}, \bibinfo {author} {\bibfnamefont {O.}~\bibnamefont
  {Brzobohat{\'y}}}, \bibinfo {author} {\bibfnamefont {K.}~\bibnamefont
  {Dholakia}}, \ and\ \bibinfo {author} {\bibfnamefont {P.}~\bibnamefont
  {Zem{\'a}nek}},\ }\href@noop {} {\bibfield  {journal} {\bibinfo  {journal}
  {Laser Phys. Lett.}\ }\textbf {\bibinfo {volume} {8}},\ \bibinfo {pages} {50}
  (\bibinfo {year} {2011})}\BibitemShut {NoStop}%
\bibitem [{\citenamefont {Zem{\' a}nek}\ \emph {et~al.}(2016)\citenamefont
  {Zem{\' a}nek}, \citenamefont {{\v S}iler}, \citenamefont {Brzobohat{\' y}},
  \citenamefont {J{\' a}kl},\ and\ \citenamefont {Filip}}]{ZemanekJOpt16}%
  \BibitemOpen
  \bibfield  {author} {\bibinfo {author} {\bibfnamefont {P.}~\bibnamefont
  {Zem{\' a}nek}}, \bibinfo {author} {\bibfnamefont {M.}~\bibnamefont {{\v
  S}iler}}, \bibinfo {author} {\bibfnamefont {O.}~\bibnamefont {Brzobohat{\'
  y}}}, \bibinfo {author} {\bibfnamefont {P.}~\bibnamefont {J{\' a}kl}}, \ and\
  \bibinfo {author} {\bibfnamefont {R.}~\bibnamefont {Filip}},\ }\href
  {http://stacks.iop.org/2040-8986/18/i=6/a=065402} {\bibfield  {journal}
  {\bibinfo  {journal} {J. Opt.}\ }\textbf {\bibinfo {volume} {18}},\ \bibinfo
  {pages} {065402} (\bibinfo {year} {2016})}\BibitemShut {NoStop}%
\bibitem [{\citenamefont {Redner}(2001)}]{bookRedner}%
  \BibitemOpen
  \bibfield  {author} {\bibinfo {author} {\bibfnamefont {S.}~\bibnamefont
  {Redner}},\ }\href {\doibase 10.1017/CBO9780511606014} {\emph {\bibinfo
  {title} {A guide to first-passage processes}}}\ (\bibinfo  {publisher}
  {Cambridge University Press},\ \bibinfo {year} {2001})\BibitemShut {NoStop}%
\bibitem [{\citenamefont {Risken}(1996)}]{RiskenBook}%
  \BibitemOpen
  \bibfield  {author} {\bibinfo {author} {\bibfnamefont {H.}~\bibnamefont
  {Risken}},\ }\href@noop {} {\emph {\bibinfo {title} {The Fokker-Planck
  Equation: Methods of Solutions and Applications}}},\ \bibinfo {edition}
  {2nd}\ ed.,\ Springer Series in Synergetics\ (\bibinfo  {publisher}
  {Springer},\ \bibinfo {year} {1996})\BibitemShut {NoStop}%
\bibitem [{\citenamefont {Seifert}(2012)}]{Seifert2012}%
  \BibitemOpen
  \bibfield  {author} {\bibinfo {author} {\bibfnamefont {U.}~\bibnamefont
  {Seifert}},\ }\href {http://stacks.iop.org/0034-4885/75/i=12/a=126001}
  {\bibfield  {journal} {\bibinfo  {journal} {Rep. Prog. Phys.}\ }\textbf
  {\bibinfo {volume} {75}},\ \bibinfo {pages} {126001} (\bibinfo {year}
  {2012})}\BibitemShut {NoStop}%
\bibitem [{\citenamefont {Mart\'{\i}nez}\ \emph {et~al.}(2015)\citenamefont
  {Mart\'{\i}nez}, \citenamefont {Rold\'an}, \citenamefont {Dinis},
  \citenamefont {Petrov},\ and\ \citenamefont {Rica}}]{MartinezPRL2015}%
  \BibitemOpen
  \bibfield  {author} {\bibinfo {author} {\bibfnamefont {I.~A.}\ \bibnamefont
  {Mart\'{\i}nez}}, \bibinfo {author} {\bibfnamefont {E.}~\bibnamefont
  {Rold\'an}}, \bibinfo {author} {\bibfnamefont {L.}~\bibnamefont {Dinis}},
  \bibinfo {author} {\bibfnamefont {D.}~\bibnamefont {Petrov}}, \ and\ \bibinfo
  {author} {\bibfnamefont {R.~A.}\ \bibnamefont {Rica}},\ }\href {\doibase
  10.1103/PhysRevLett.114.120601} {\bibfield  {journal} {\bibinfo  {journal}
  {Phys. Rev. Lett.}\ }\textbf {\bibinfo {volume} {114}},\ \bibinfo {pages}
  {120601} (\bibinfo {year} {2015})}\BibitemShut {NoStop}%
\bibitem [{\citenamefont {Bodrova}\ \emph {et~al.}(2016)\citenamefont
  {Bodrova}, \citenamefont {Chechkin}, \citenamefont {Cherstvy}, \citenamefont
  {Safdari}, \citenamefont {Sokolov},\ and\ \citenamefont
  {Metzler}}]{Bodrova2016}%
  \BibitemOpen
  \bibfield  {author} {\bibinfo {author} {\bibfnamefont {A.~S.}\ \bibnamefont
  {Bodrova}}, \bibinfo {author} {\bibfnamefont {A.~V.}\ \bibnamefont
  {Chechkin}}, \bibinfo {author} {\bibfnamefont {A.~G.}\ \bibnamefont
  {Cherstvy}}, \bibinfo {author} {\bibfnamefont {H.}~\bibnamefont {Safdari}},
  \bibinfo {author} {\bibfnamefont {I.~M.}\ \bibnamefont {Sokolov}}, \ and\
  \bibinfo {author} {\bibfnamefont {R.}~\bibnamefont {Metzler}},\ }\href
  {\doibase 10.1038/srep30520} {\bibfield  {journal} {\bibinfo  {journal} {Sci.
  Rep.}\ }\textbf {\bibinfo {volume} {6}},\ \bibinfo {pages} {30520} (\bibinfo
  {year} {2016})}\BibitemShut {NoStop}%
\bibitem [{\citenamefont {Arold}\ \emph {et~al.}(2018)\citenamefont {Arold},
  \citenamefont {Dechant},\ and\ \citenamefont {Lutz}}]{ArnoldPRE2018}%
  \BibitemOpen
  \bibfield  {author} {\bibinfo {author} {\bibfnamefont {D.}~\bibnamefont
  {Arold}}, \bibinfo {author} {\bibfnamefont {A.}~\bibnamefont {Dechant}}, \
  and\ \bibinfo {author} {\bibfnamefont {E.}~\bibnamefont {Lutz}},\ }\href
  {\doibase 10.1103/PhysRevE.97.022131} {\bibfield  {journal} {\bibinfo
  {journal} {Phys. Rev. E}\ }\textbf {\bibinfo {volume} {97}},\ \bibinfo
  {pages} {022131} (\bibinfo {year} {2018})}\BibitemShut {NoStop}%
\bibitem [{\citenamefont {Lamouroux}\ and\ \citenamefont
  {Lehnertz}(2009)}]{LamourouxPLA09}%
  \BibitemOpen
  \bibfield  {author} {\bibinfo {author} {\bibfnamefont {D.}~\bibnamefont
  {Lamouroux}}\ and\ \bibinfo {author} {\bibfnamefont {K.}~\bibnamefont
  {Lehnertz}},\ }\href {\doibase 10.1016/j.physleta.2009.07.073} {\bibfield
  {journal} {\bibinfo  {journal} {Physics Letters A}\ }\textbf {\bibinfo
  {volume} {373}},\ \bibinfo {pages} {3507 } (\bibinfo {year}
  {2009})}\BibitemShut {NoStop}%
\bibitem [{\citenamefont {O'Hagan}\ and\ \citenamefont
  {Leonard}(1976)}]{OHaganBiometrika76}%
  \BibitemOpen
  \bibfield  {author} {\bibinfo {author} {\bibfnamefont {A.}~\bibnamefont
  {O'Hagan}}\ and\ \bibinfo {author} {\bibfnamefont {T.}~\bibnamefont
  {Leonard}},\ }\href {\doibase 10.2307/2335105} {\bibfield  {journal}
  {\bibinfo  {journal} {Biometrika}\ }\textbf {\bibinfo {volume} {63}},\
  \bibinfo {pages} {201} (\bibinfo {year} {1976})}\BibitemShut {NoStop}%
\bibitem [{\citenamefont {Tolman}(1938)}]{TolmanBook}%
  \BibitemOpen
  \bibfield  {author} {\bibinfo {author} {\bibfnamefont {R.~C.}\ \bibnamefont
  {Tolman}},\ }\href@noop {} {\emph {\bibinfo {title} {The Principles of
  Statistical Mechanics}}}\ (\bibinfo  {publisher} {Clarendon Press},\ \bibinfo
  {year} {1938})\BibitemShut {NoStop}%
\bibitem [{\citenamefont {Tolman}(1918)}]{Tolman1918}%
  \BibitemOpen
  \bibfield  {author} {\bibinfo {author} {\bibfnamefont {R.~C.}\ \bibnamefont
  {Tolman}},\ }\href {\doibase 10.1103/PhysRev.11.261} {\bibfield  {journal}
  {\bibinfo  {journal} {Phys. Rev.}\ }\textbf {\bibinfo {volume} {11}},\
  \bibinfo {pages} {261} (\bibinfo {year} {1918})}\BibitemShut {NoStop}%
\bibitem [{\citenamefont {Huang}(1987)}]{HuangBOOK}%
  \BibitemOpen
  \bibfield  {author} {\bibinfo {author} {\bibfnamefont {K.}~\bibnamefont
  {Huang}},\ }\href@noop {} {\emph {\bibinfo {title} {Statistical
  Mechanics}}},\ \bibinfo {edition} {2nd}\ ed.\ (\bibinfo  {publisher}
  {Wiley},\ \bibinfo {year} {1987})\BibitemShut {NoStop}%
\bibitem [{\citenamefont {Sekimoto}(2010)}]{SekimotoBOOK}%
  \BibitemOpen
  \bibfield  {author} {\bibinfo {author} {\bibfnamefont {K.}~\bibnamefont
  {Sekimoto}},\ }\href {\doibase 10.1007/978-3-642-05411-2} {\emph {\bibinfo
  {title} {Stochastic Energetics}}}\ (\bibinfo  {publisher} {Springer-Verlag
  Berlin Heidelberg},\ \bibinfo {year} {2010})\BibitemShut {NoStop}%
\end{thebibliography}%

\end{document}